# Rock Joint Surfaces Measurement and Analysis of Aperture Distribution under Different Normal and Shear Loading Using GIS

MOSTAFA SHARIFZADEH[*,1,2], YASUHIRO MITANI[1], TETSURO ESAKI[1]

[1]*Institute of Environmental Systems, Faculty of Engineering, Kyushu University, Japan*

[2]*Department of Mining, Metallurgy and Petroleum Engineering, Amirkabir University of Technology, Tehran, Iran*

## Summary

Geometry of the rock joint is a governing factor for joint mechanical and hydraulic behavior. A new method of evaluating aperture distribution based on measurement of joint surfaces and three dimensional characteristics of each surface is developed. This method allows us to determine and visualize aperture distribution under different normal stresses and shear displacements, which is difficult to observe experimentally. A new laser scanner system is designed and developed for joint surface measurements. Special attention is paid to both surfaces' data gained by measurements and processing, such as x-y coordinate table modification, data referencing and matching between upper and lower surfaces. Artificial joint of granite surfaces are measured, processed, analyzed and three dimensional approaches are carried out for surface characterization. Parameters such as "asperity's heights", "slope angles", and "aspects" distribution at micro scale, local concentration of elements and their spatial localization at local scale are determined by Geographic Information System (GIS). These parameters are used for joint's surfaces matching and its real behavior quantitative analysis. The upper surface is brought down to make contact with the lower surface and distance between the two surfaces is obtained from the joint mean experimental aperture, which is acquired from normal and shear tests. Changes of aperture distribution at different normal stresses and various shear displacements are visualized and interpreted. Increasing normal load causes negative changes in aperture frequency distribution which indicates high joint matching. However, increasing shear displacement causes a rapid increase in the aperture and positive changes in the aperture frequency distribution which could be due to unmatching, surface anisotropy and spatial localization of contact points with proceeding shear.

*Key words:* rock joint, laser scanner, joint surfaces morphology, GIS, aperture distribution, aperture visualization, joint shear

# 1. Introduction

Design and construction of deep underground structures utilize rock mass characteristics such as high-rigidity, sealing, durability and isolation. More severe design conditions and more accurate properties of rock mass are needed in such a deep underground development from safety, economical and environmental points of view. Most of the latest proposed underground developments employ sealing and isolation property of rock mass. It is important to examine the permeability of rock mass wherein underground structures are to be constructed, in order to confirm its capacity to isolate. The first step in understanding rock mass conductivity is the comprehension of single rock joint conductivity. The hydraulic conductivity and mechanical behavior of the joint is functions of its surface morphology as well as aperture distribution.

Several techniques have been used for aperture measurement which could be categorized in two main groups of direct and indirect methods. Direct methods include injection, x-ray computer tomography (CT) and nuclear magnetic resonance imaging (NMRI). Indirect methods use joint surfaces data obtained from touch-type mechanical profilometer, laser scanner or photogrametry to calculate the joint aperture distribution. Although the proposed aperture measurement methods have improved our understanding of aperture determination, a precise method to evaluate aperture distribution under different normal and shear loading processes is still required.

Direct aperture measurement methods (Gale *et al*. (1987, 1990), Hakami (1992, 1995)) uses resin injection technique. Pyrak-nolte *et al*. (1987) uses wood metal injection under normal stress to evaluate joint aperture. Gentier *et al*. (1989) and Hakami (1992) made transparent replica of void spaces of joint and determined aperture, through image analysis of sections. Making aperture and surface replica is known as destructive methods where the use of the same specimen for aperture determination and experimental test is impossible. In addition, these methods have low accuracy because injected material's viscosity prevents filling of small voids. Furthermore, thick cutting intervals provide only a little data of aperture. Kumar *et al.* (1997) and Dijk *et al.* (1999) used NMRI to measure the aperture distribution. Johns *et al.* (1993), Keller (1998), Ohtani *et al.* (2000) and Stephanie *et al.* (2001) used the X-ray computerized tomography (CT) technique for aperture measurement. The X-ray CT and NMRI methods have low spatial resolution, high cost and difficulties with calibration, measurement and analysis. Indirect aperture measurement methods (Brown *et al.* (1985) and Gentier (1986)) measured joint surfaces topography by mechanical touch-type profilometer. For each surface, surface height and location is continuously recorded. The upper surface is brought to opposing surface and aperture is computed by knowing the distance between the joint surfaces. Measuring topography, by mechanical touch-type profilometer has low accuracy due to wide measurement intervals and is not able to measure sharp points. This method could also damage the surface. Furthermore, there is difficulty in referencing and matching between two surfaces. Esaki *et al.* (1995), Iwano *et al.* (1995) introduced laser scanner instead of mechanical profilometer to overcome these difficulties. They succeeded to measure surface with very small intervals about 0.5 – 1.0 mm without specimen damage, also they used special marked points to reference the upper surface to the lower one. Later Lee *et al.* (2002) also used similar method for aperture determination. Hans *et al.* (2003) used laser scanner for joint surfaces measurement at each defined shear displacement during the test and calculated the aperture using the surface data. All aperture measurement techniques have advantages and limitations that must be chosen depending on the study.

Joint aperture determination using asperity data completely dependent upon joint surface characteristics and comprehensive surface roughness characterization which can lead to correct aperture determination. Roughness can be explained as local departures from planarity and many researchers so far, have presented two dimensional characteristics of surface roughness using profiles. However, surface irregularity, heterogeneity and anisotropy can not be depicted in two dimensions. Statistical (Tse and Cruden (1979)), geostatistical (Gentier *et al*. (2000)) and fractal (Kultilake *et al*. (1995), Xie *et al*. (1997), Lanaro *et al*. (2000) and Fardin *et al*. (2001)) analysis are also used to characterize surface morphology; however, one major problem is inability of reproducibility, where surfaces with similar statistical or fractal values show different spatial distributions. Even though these approaches are useful to improve our understanding of roughness, they are not sufficient for three dimensional roughness characterization. Surface roughness is spatially localized and it is necessary to characterize the surface using three dimensional characteristics. Several attempts have been made to present three dimensional characteristics of a surface. Gentier *et al.* (1997), Lanaro *et al.* (2000), Belem *et al*. (2000), Grasselli *et al*. (2000, 2003) and Sharifzadeh *et al.* (2004a) are among the few researchers who have taken into consideration the three dimensional characteristics of joint's surfaces such as surface aspect and anisotropy to explain mechanical behavior of a joint.

Research shows that aperture is dependent upon stress history, normal displacement, shear displacement and scale of study. Even though aperture measurement and its distribution analysis have been conducted by previous researchers, yet above problems remain unsolved. Hydromechanical tests provide the only result for whole test specimen. However, the micro-mechanism of aperture and contact distribution, asperity deformation and flow test process inside the joint is not clear. This study aims to determine, visualize and interprete aperture distribution under different normal stresses history and various shear displacements in three dimensions.

## 2. Procedure for Determining the Aperture Distribution

The procedure to achieve precise aperture distribution using joint surfaces data is illustrated in Fig. 1. The application of originally designed and developed high resolution laser scanner is justified for joint surfaces asperity height measurement. Errors in surface data originated from laser measurement system and specimen installation on x-y positioning table are rectified by this method. Upper and lower surfaces are characterized and their comparison is used to confirm matching of the joint surfaces. Moreover, normal loading test to determine initial aperture at different normal stresses and direct shear test to determine aperture changes during shear is performed. Finally, joint surfaces and aperture data are integrated in GIS to determine aperture distribution. Aperture distribution is determined and visualized under different stresses and displacement conditions. Variation of aperture distribution during normal loading and shear process are quantified by using both aperture distribution map and aperture frequency distribution.

## 3. Joint Surface Measurement and Characterization

Rock joint surface roughness plays a major role in joint mechanical and hydraulic behavior. Therefore, it is clear that a precise measurement of rough surface topography is a key to understand the joint mechanical and hydraulic behavior specially during shearing. It is obvious that the contribution of roughness in hydromechanical models strongly depends on the method of surface measurement and quantification. For more accurate measurements of joint surfaces morphology, a new device using laser beam has been designed and developed (Fig. 2-a,b). Surface data were measured on a virtual grid mesh with square elements sized 0.2mm (Fig. 2c). The measurement area was set wider than specimen surface area. Then GIS was used to extract real surface data, and to check several possible errors and correct them. Finally, both surfaces' three dimensional characteristics were determined using GIS.

### 3.1 Laser Scanning Apparatus and Measurement Procedure

To obtain precise measurement of asperity's height, a new device using a laser beam was designed and calibrated (Fig. 2a). The new device consists of: 1) laser displacement sensor head, with resolution of 0.5μm, spot size of 45×20μm and measurement height range of ±8mm, 2) x-y positioning table with stroke area of 250×150mm having positioning accuracy of ±15μm and repositioning accuracy of ±3μm, 3) multi-servo controller of x-y table, 4) laser displacement sensor head controller, and 5) data recording and control devices. In this system, the distance between the rough surface and the laser gauge is measured as z-elevation by measuring head. The zero reference position and the intervals of measurement in x- and y- directions are controlled by positioning table controller. The asperity data are first measured in x-direction then the table moves in y-directions with defined intervals. The process is repeated until the whole surface measurement is completed. This system has the capability of setting area and intervals of measurement in arbitrary values.

One of the most important difficulties of this technique (surface topography) - especially for joint aperture determination - is its high sensitivity to referencing the specimen (the upper and lower halves with respect to each other) and matching the joint surfaces based on their measured asperity height. To overcome this problem, both experimental surface measurement method and surface data analysis method are improved to achieve matching between the two surfaces (Sharifzadeh *et al*. 2004b). As shown in Fig. 2b, a positioning table is equipped by a so called "Setting Block" with micrometers and pins to fix the specimen on the table to prevent possible specimen displacements during measurement. Four sides of the specimen are kept tight where the Setting Block holds at two sides, the other two sides are kept by placing two pins at each side situated beside the micrometers (Fig. 2b). Micrometers perform double duty before and after surface measurement. The specimen position is measured by two micrometers at each side in order to compare whether the repositioning of the specimen is achieved or not.

The results of micrometer measurements are used to set the other half of the specimen at the same position of the previous measured half.

In this study, a granite block of 180mm×100mm×80mm (length × width × height) was used. An artificial rock joint is used in order to raise the reproducibility. An artificial joint was created parallel to the sedimental plane at mid-height of the specimen. An apparatus was made for creating joint as detailed by Esaki *et al*. (1998). The apparatus consists of a couple of horizontal jacks and a normal loading jack. The horizontal jacks are attached to the long side of specimen for creating joint by splitting with a couple of steel wedges in order to control the aperture of the joint. For stability in creation of the joint, a constant horizontal load for splitting is applied through pair of wedges after applying a prescribed normal load on the specimen. Then, the normal load is gradually reduced during fracturing while horizontal load is kept constant. Thus, joint is made stable under controlled conditions and can be used for surface measurement. The procedure for surface measurement consist of: i) placing the specimen upper or lower surfaces on x-y positioning table, ii) tightening long side and then short side pins and measuring long side and short side micrometers, iii) setting the coordinate of starting point for measurement, iv) inputting parameters into PC (starting point, number of lines and intervals in x- and y- directions) and finally v) running the code. Our code records x-y table coordinates and the correspondent of z- values measured by laser scanner.

### 3.2 Laser Scanner Calibration

Several parameters are important for accurate measurement, such as; i) temperature, ii) average measurement frequency, iii) table stopping time for laser radiation – reflection, iv) installation of laser displacement meter, and v) x-y positioning table movement speed. These parameters can be fixed based on the rock conditions and measurements. Efforts were made to isolate the changes of temperature and light intensity during day and night. However, errors due to small changes are often unavoidable. The calibration of specimen setting and x-y positioning table speed for laser measurement is explained below:

A 40mm×20mm area in the center of the specimen is selected and several measurements in the same condition were repeated with respect to long side and short side setting and with different table speeds. For setting up the specimen on the table and giving the priority of tightening pins at the long side or short side, five measurements for each setting were repeated and the standard deviation of the settings is determined. The standard deviation with respect to the long side is 3.4E-02 in μm and the short side is 3.6E-02 in μm. Since long side standard deviation is less than that of the short side, the long side setting is to be made prior to short side setting. In other words, at first it is better to tight long side pins and then tight the short side pins. The effects of the table speed, stopping time for each laser measurement and loop number on the accuracy of laser scanner measurement results were checked with different speeds and standard deviations of measurements. It is noteworthy that loop number is the representative of table speed and low loop number, indicates slow movement or long stopping time of table and vice versa. Results show the same standard deviation up to 300 loops and after that a gradual increases. Therefore, 300 loops is chosen for measuring (Fig. 3). This system is used for precise measurement of joint surfaces mesh element height.

### 3.3 Joint's surfaces Data Processing

GIS technology along with specifically developed computer programs were implemented to measure asperity heights, processed data and evaluate possible deviations and inclination correction to obtain spatial positioning and matching of joint surfaces. A virtual mesh having a square element size of 0.2mm spread on each surface and each element height is measured by laser scanner (Fig. 2.c). It should be noted that, based on surface asperity conditions, each element could be an asperity or several elements may form one asperity. In other words selected measurement intervals or element size is small enough to represent asperity. Thus, in some cases element could be used as asperity or several elements with each other could form asperity. Each half is placed on the laser scanner table and surface element height was measured with 0.2mm intervals in x- and y- directions. Measured surface data are checked against possible installation and intrinsic laser measurement errors. Possible errors originated from specimen installation, such as surface deviation from x-y plane and inclination from x-z and y-z planes are checked as shown in Fig. 4. The procedure for surface deviation check is illustrated in Fig. 5. To avoid loosing surface data, the measurement area was set 1mm wider (182×102) compared to the original size. A total of 464,100 elevation data are obtained on each surface with their x- and y- coordinates. The Geographic Information System (GIS) is used to visualize and define real surface mesh element height data from extra error point data. Surface data are classified in different height classes. GIS-selection tools are used to find the border of real rock joint surface from adjacent error points (Fig. 5a). Real surface border lines are determined by linear fitting of selected border points (Fig. 5b). Border lines equations are used to calculate the joint center ($x_c$, $y_c$) and the angle of surface long side line from x- axis (Fig. 5c). By comparing the long side border lines, equations of both surfaces deviated at x-y plane and difference in border lines deviation are calculated. The deviated surface data are rotated back to fit on each other. In this study, the difference between upper and lower surface borderline angle is α = 0.145°, which is corrected by rotating upper surface data with respect to the center of the specimen. In order to check the x-z and y-z plane inclinations (Fig. 4c), the surface equation for each surface are calculated using element height data. Equations for both lower and upper surface are compared to each other to check the parallelism and inclination of surfaces as shown in Fig. 4c. In this study, the upper and lower surfaces equations are determined as;

$$\begin{cases} \textit{Lower surface equation}: & z = -0.00005x + 0.00017y - 1.378 \\ \textit{Upper surface equation}: & z = -0.00005x + 0.00022y + 0.895 \end{cases} \quad (1)$$

The equation (1) shows the joint surface equation for the presented sample. Comparison of the lower and upper surface equations shows close similarity between two surfaces, hence joint surfaces are assumed to be parallel. Finally, data of joint surfaces with a total of 449,599 heights and over 180mm× 99.8mm area were extracted.

Intrinsic errors from laser measurement are due to the differences of laser light reflection from dark and bright minerals on surfaces. Errors occurred from laser measurement noise give an out of scale value (±99.99 in mm) and abnormal z-value (larger than five times the data standard deviation) in some points. In this case there were 20 points (12 points in lower and 8 points in upper surface) with z- value of ±99.99 in mm and 83 points (51 points in lower and 32 points in upper surface) with z- value greater than five times of standard deviation. The error points in both upper and lower surfaces are corrected by using the average of neighbor point's elevations. Since joint upper surface is turned back to set on laser scanner table. After obtaining the real surface, upper surface data is numerically overturned to represent its natural condition in joint. The result the modified surface mesh element heights of both upper and lower surfaces data is saved as ASCII Grid and xyz format to be used for joint aperture assessment.

### 3.4 Joint Surface Characterization

To evaluate the surface data and modification methods discussed in previous section and verify the capability of referencing joint surfaces, upper and lower surfaces characteristics are determined and compared with each other and check whether they match or do not. Most researchers so far have focused on sectional, i.e., two-dimensional simulation of surface roughness although the actual morphology is three-dimensional. Basically, surface roughness is most often described by some linear parameters calculated in individual profiles which are obtained along one-directional parallel lines (Belem *et al.* 2000). The parameters describing entire surface of the fracture can be obtained by calculating the average of parameters of all profiles (Barton *et al.* 1982). Hence, a three dimensional problem is solved by two dimensional approaches. However, real geometry should be quantified by three dimensional characterizations of the surfaces. Surface asperities have different characteristics: they are not evenly distributed and their distribution depends on spatial position of the surfaces. In this study, efforts have been made to give three-dimensional interpretations of joint surface roughness and aperture distribution based on the height of local roughness elements using GIS and statistical parameters. The data of both upper and lower surfaces mesh elements converted to ASCII grid as GIS "raster" data. Each 0.2mm×0.2mm area is measured as one element or point data in mesh on surfaces mesh, and collection of elements form asperities on joint surfaces. Consequently, each surface is defined as a collection of asperities with different heights, slope angles, aspects and positions. Using GIS three dimensional analysis tools, these characteristics could be analysed. Each surface is studied with three scales consisting: i) element or micro-scale, ii) local area scale (or a few square centimeters), and iii) laboratory scale. Micro-scale characteristics are: elements heights, slope angles and aspects determined accurately by using GIS (Fig. 6); Elements concentrations on local area on joint surfaces are presented as local scale characterization. Finally, micro and local scale results are used to illustrate whole surfaces characteristics. Therefore, three important micro-scale characteristics of joint surfaces are illustrated and their distribution is determined.

The first property is element's height. To obtain the height of each element mean surface data are calculated as base level and elements height is measured with respect to mean surface. Thus element height represents the height from surface average height. Since the upper and lower surface is similar their mean data is almost similar. Element heights form asperities and joint roughness and pattern of asperities distribution on the surfaces through which the governs both mechanical and hydraulic behaviors. Fig. 6a,d shows the asperity's heights distribution maps and frequency distribution histograms with statistical calculation results. The lower and upper surfaces asperity distribution maps (Figs. 6a,b) and their histograms (Figs. 6c,d), show normal distribution with mean height from standard line, and standard deviation of 0.7448 mm and 0.7550 mm, respectively.

Micro and local scale surface Comparison shows that normal distribution is found in micro scale as shown in Fig. 6c,d and spatially localized distribution is observed in local scale (Fig. 7a,b) which indicates irregularity of asperities distribution. i.e., although normal distribution of elements' heights for whole surface is shown in Fig. 6c,d, - and it is expected a regular asperities distribution,- (Fig. 7a,b) irregular changes of asperities is found in the profiles (Spatial localization).

Surface mesh element plane angularity which is the element plane inclination angle with respect to horizontal plane will be referred to element "slope angle" here after. Slope angle is a direct measure of joint surfaces matching and dilation during shear process. Figs. 6e,f shows slope angle frequency distribution histograms. Both histograms show log-normal distribution, having mean slope angles of $22.7°$ and $24.1°$, and standard deviation of $12.6°$ and $13.0°$, respectively. Surface asperity plane orientation referred to as "aspect" hereafter is one of the most important characteristics affecting joint shear behavior. "Aspect is the down slope direction of an element to its neighbor elements and could be identify the orientation or direction of slope. "Aspect" can be defined as the angle between normal vectors on each triangle from north (shear) direction, which is a projection of plane orientation with respect to shear direction (Grasselli *et al*. 2003). Fig. 6g,h show polar plot of aspect's direction that indicates equal distribution of aspect over joint surfaces. Thus, equally dilation is likely to occur in all shear directions.

Joint upper and lower surface characteristics are compared to each other to evaluate the capability of surface morphology measurement and data modification method. Comparison between pairs of upper and lower surface asperity heights, slope angles and aspects direction distribution as shown in Figs. 6c,d, e,f, g,h indicate close similarity between two surfaces which means high matching of joint surfaces. Thus, it could be concluded that the applied data measurement technique (surface topography) is successful to overcome referencing difficulties which encountered in former researches and achieve matching between joint surfaces sing surface topography method. On the other hand, the referencing is obtained in surface topography method.

From all of the above mentioned parameters, a three dimensional interpretation of joint surfaces is presented. This analysis can illustrate the aperture distribution during the shear. Each surface can be defined as a collection of asperities with different heights, slopes, aspects and statistical values. The surface contains a collection of elevated and depressed elements having their own statistical quantities (Fig. 7), which are different from each other. The difference between concentration parameters and their spatial distribution verfies surface irregularity which can be seen in Fig. 7. On the other hand, if the changes of asperity height trace in line will give two dimensional profiles, which show roughness, waviness and undulation. However, in case of three dimensional characterizations, we assumed spatial distribution of asperities concentration with different characteristics. Therefore, the joint surface behavior not only depends on micro scale characterizations, but also on local scale characterizations of asperity concentrations which is spatially distributed unevenly on the surfaces, thus, it plays a major role on the shape of joint surface and govern the joint aperture during shear displacement.

### 4. Initial Mechanical Aperture Under Normal and Shear Loading

To establish article aperture, joint surfaces should be closed to each other. To bring the two surfaces together, the distance between them should be determined. In this study, the distance is calculated from initial aperture and dilation data which are obtained from experiments. These apertures are compared with mean aperture during aperture distribution calculation using the joint asperity height method. Therefore, distance between two surfaces is a very important parameter in joint aperture distribution determination and must be calculated precisely. In this study great attention is paied to precise determination of aperture under normal loading and shear displacements.

The initial aperture under different normal stresses is determined. To do this, a normal loading test is carried out on the joint. Normal displacement is measured by four transducers at the four edges of the specimen upper box. Loading of 10MPa and unloading of 1MPa operation is repeated few times until a stable curve is obtained (Fig. 8a: curve-a). Normal displacement curve includes the deformation of the joint itself, intact part of rock specimen and the shear box. There are no normal displacement data at low normal stresses (dashed line in curve-a in Fig. 8a.) because of the weight of upper shear box, thus the intersection with horizontal axis is unknown. To calculate the intercept of curve-a, the values for normal stresses for 3MPa is selected and fitted with hyperbolic function. Therefore, interception on horizontal axis is determined and curve-a shifted to origin (Fig. 8a: curve-$a_0$). As applied, normal stress is increased, and the normal stiffness remains almost constant and represents the normal stiffness of the intact rock and the shear apparatus. This stiffness is assumed to remain constant during the test. Intact rock and apparatus stiffness is calculated from constant part of curve-a. To do this, the data larger than 8MPa is selected and fitted by a line (Fig. 8a: curve-b) and similarly curve-b is shifted to the origin (Fig. 8a: curve-$b_0$). After translating curve-a and curve-b to the origin, the normal deformation curve of the rock joint is obtained (Fig. 8b). The closure is approximated by a hyperbolic function of Bandis *et al*. (1983) as;

$$\sigma_n = \frac{u_n}{a - bu_n} \quad (2)$$

Therefore maximum closure $V_m$ is obtained as follow;

$$V_m = \lim_{\sigma_n \to \infty} u_n = \frac{a}{b} \tag{3}$$

where $\sigma_n$ is the normal stress, $u_n$ is the normal deformation, $V_m$ is the joint maximum closure, $a$, $b$ are sample coefficients. Initial aperture is calculated by taking the difference of normal deformation from the maximum closure. The joint aperture in proportion to normal stress is obtained as follow (Fig. 8c).

$$e_m = V_m - u_n = \frac{V_m}{1 + b\sigma_n} \tag{4}$$

Where $e_m$ is the initial mechanical aperture or distance between the two surfaces at rest (in mm) and $\sigma_n$ is the normal stress. For this study, equation (4) becomes as follow (Fig. 9):

$$e_m = \frac{0.099}{1 + 1.14\sigma_n} \tag{5}$$

The initial aperture varies from 0.0607mm at 1MPa to 0.0087mm at 10MPa. Thus, with increasing normal stress up to 4MPa aperture decreases rapidly, but after it follows a gradual decrease up to 10MPa. This tend indicates that with increasing deformation and contact area, it becomes more difficult to obtain deformation.

The change of aperture at different shear displacements was obtained from joint dilation during direct shear test. Direct shear test was performed on artificial granite joint under 3MPa of normal stress (Fig. 10a). Moreover, during shear process the upper and lower surfaces are kept parallel by adjusting normal load on the front and rear normal jacks.

Prior to shear, it is noteworthy that the uniaxial compression test is also carried out on granite intact specimen. The compressive strength of the rock material was 165MPa which specifies it a very hard rock. Normal stress during shear is very low compared to the rock strength. Thus, it is assumed that under this condition, gouge material will not be produced and asperities will behave as rigid material. No asperity damage is considered in this research.

Changes of shear stress and normal displacement with respect to shear displacement are shown in Fig. 10a,b. The normal displacement before shear (at zero shear displacement) is obtained from joint initial aperture curve as in section 4. Shear displacement proceeded up to 20mm, and the changes of normal displacement during shear is shown in Fig. 10b. Small contractions is observed up to 1.5mm of shear because of locking of the asperities of upper and lower surfaces. This is followed by a sudden increase in normal displacement (dilation), simultaneously shear stress increases up to peak. After that, shear stress rapidly decreases to residual stress and becomes almost constant (Fig. 10a). The changes of normal displacement during shear were used to determine the aperture at different shear displacements.

As a result, the distance between the joint surfaces for aperture distribution analysis is determined from initial aperture according to normal stresses and changes of aperture at different shear displacement. These data will be used as a part of input data for aperture distribution analysis.

## 5. Aperture Distribution at Normal and Shear Loading

To obtain the aperture distribution using surface asperity's heights data, two halves are brought to proximity with each other. For this purpose, the lower surface is kept fixed and upper surface is brought down to form aperture. By assuming that two surfaces are coincided (point by point), the aperture can be defined as the distance or gap between two corresponding elements on both joint's surfaces mesh.

Joint aperture distribution is determined using newly defined surface characteristics, integrated in GIS (Sharifzadeh 2005). This can be performed by importing the upper and lower surfaces x-, y- and z- data in GIS. By using GIS tools, the information on two surfaces can be calculated and new data tables are created. This data table consists of four columns: the x- y- coordinate and z-elevation of the upper and lower surfaces simplified as the same x-y having two z-elevation values. With same x-y- grid and different z- elevations new data creation becomes possible. New columns showing the piecewise aperture is created and the mean value of each column is calculated and compared with aperture from experiment (obtained aperture from previous section). When the mean column value becomes equal to the initial aperture, calculations is stopped and results are the aperture value at specific case. During shear (according to real shear testing machine) lower surface grid data are moved left with defined steps to make shear. This procedure is repeated until the completion of all columns for different normal stresses and shear displacement conditions. As well as aperture distribution was visualized in GIS using these data and statistical parameters were obtained. Elements with negative and null values represent elements in contact or compression and assumed as contact points. Elements with positive values represent the aperture. Mean aperture and contact ratio are calculated and also data table is used to show aperture frequency distribution histograms.

## 5.1 Aperture Distribution at Different Normal Stresses

Aperture distribution varies with normal stresses and can be calculated and visualized using the above mentioned method. Fig. 11 shows the aperture distribution map (left), its frequency distribution, the mean aperture and contact ratio at 1, 3, 5 and 10MPa of normal stresses. In Fig. 11a,d, contact ratio increases from 84.8 to 98.4 percent, while mean aperture decreases from 43.97 to 7.29 microns for normal stresses increasing from 1to 10 MPa, which shows high rate of matching between two surfaces and approves our surface measurement and processing technique and specify the correct aperture determination method.

The frequency distribution histogram of the aperture is presented in Fig. 11 (right). Both surfaces asperities heights follow Gaussian distributions. However, the aperture frequency distribution under normal load is found similar to Poison or Log-normal distribution. Increasing of normal stresses causes drop in aperture, and increase in the contact ratio, thus results to negative change (shifting to left) in aperture frequency distribution indicating that whole surfaces come to close contact and therefore high matching between two surfaces is found. At high normal stresses, contact ratio is almost 100% and joint is completely closed which indicate that joint behaves as intact rock. However, a small value of aperture is still remaining.

5.2 Aperture Distribution at Different Shear Displacements

Changes of aperture distribution during shear are illustrated by a three dimensional surface characterization based on results of shear experiments. Fig. 11b shows aperture distribution map (left) and frequency distribution (right) before shear and Figs. 12a-h show the change of aperture distribution map (left) and aperture frequency distribution (right) during different shear displacements. In this study the effect of the gauge material is neglected. Comparison of the aperture maps with aperture frequency distribution before shear shows that contact elements are distributed equally because of well matching between two surfaces (Fig. 11b). With increasing shear, at initial sliding some asperities leave contact, thus contact ratio decreased and mean aperture increased greatly, but aperture and contact pattern still show an even distribution (Fig. 12a-b). At critical point near peak shear, where the shear stress is in equilibrium with the strength of asperities in contact (Fig. 9a), an increment of shear causes simultaneous shearing of all asperities in contact. Consequently, a sudden change in aperture distribution occurs due to unmatching between two surfaces. At this stage, micro scale asperities loose contact and localized areas remain in contact as shown in Fig. 13c-e. Aperture and contact are spatially localized; as a result, the ratio of the contact area decreases and aperture increases rapidly. Finally with increasing shear displacement, dilation becomes controlled by next asperities in shear and consequently unmatching increases between the surfaces. Distribution of contact area spatially localized on certain asperities and mean aperture and contact ratio shows slight change (Fig. 12h). Aperture frequency distribution is presented in Fig. 12 (right). The contact ratio varies from 93.6% before shear to 5.1% at 20mm of shear displacement. The aperture values change from 20.79μm before shear to 1.54mm (539.96μm) at 20mm of shear displacement. Increasing shear caused an increase in aperture and a decrease in contact ratio. Following aperture frequency histograms shows that although at first aperture frequency distribution follows a Poisson distribution, with increasing shear an increase in aperture frequency which can be described with normal distribution curve occures. The results show that, for shear displacement before peak, the distribution of aperture and contact areas is evenly distributed. With increasing shear displacement, particularly in residual region, spatially localized and heterogeneous distribution of aperture is observed due to anisotropy and un-matching of surfaces.

## 6. Discussions

Aperture is very sensitive to changes of joint surface morphology and applied conditions such as changes of normal and shear load. A newly developed laser scanner is capable to measure joint surface with very small element (grid) size of 0.2mm×0.2mm. This element size is small enough to study the surface morphology. Combination of measured data in GIS is found to be a powerful tool to calculate and visualize the surface characteristics and their comparisons. In micro scale, asperity heights, slope and aspect are determined using GIS. In local scale the "elements concentration" concept is presented by aperture distribution map and used for three dimensional characterization of the surface. Each surface is described as several elements concentrations with spatial distribution on surfaces and each concentration has its own concentration (statistical) value. Based on this concept, the surface asperity frequency distribution followed the Gaussian frequency distribution. However, concentration of elements on local scale shows roughness irregularity and spatial distribution, which cause surface anisotropy and heterogeneity. Surface anisotropy and heterogeneity is the main source of joint complex behavior and unmatching with small displacement.

Furthermore, the determination of distance between two surfaces from aperture calculated through experiments make it possible to verfy aperture distribution on joint with same element size as surface measurement.

Finally, the aperture distribution is determined and visualized under different normal stresses and shear displacements using joint surface data and initial aperture integrating in GIS. Changes of aperture distribution during normal and shear process are interpreted using both aperture distribution maps and aperture frequency distribution. Aperture distribution during normal loading shows that; i) aperture is evenly distributed, ii) findings at micro-scale aperture (Fig.11) are verified at macro scale (Fig. 8 and Fig. 9), and iii) aperture decreases with increasing of normal stress and reach to residual value after specific normal stresses. Moreover, aperture distribution during shear shows that; i) before peak shear, aperture is randomly distributed; however, after peak shear, aperture distribution is spatially localized with increasing shear and depends on surface anisotropy, ii) joint surfaces anisotropy and heterogeneity is the main cause of spatial localization of aperture distribution and iii) contact ratio decreases rapidly and reaches to a quite small value at large shear displacement after peak shear stress.

It is hoped that the analysis of surface and aperture distribution at small scale would enable us to make local scale study of the hydraulic and mechanical behavior. Hence, mechanical and hydraulic properties can be specified locally for each element with different characteristics and superposition of them which represent the whole specimen behavior (Sharifzadeh, 2005).

## 7. Conclusion

In this study, efforts have been made to establish a new three dimensional joint surface measurement and characterization method to determine aperture distribution precisely and visualize it under normal and shear loading. An originally designed and developed laser scanner is capable to measure surface mesh element height with high resolution in z-elevation and very small intervals with high accuracy in x-and y-directions. An improvement in both asperity heights measurement techniques and processing could obtain referenced and matched surfaces. The joint surface processing and analysis in GIS, not only shows highly matching between two surfaces but also makes it possible for three dimensional characterizations of surfaces. Some important parameters such as asperity heights, slope angle, aspect and concentration of asperities on local areas. Although asperity heights frequency distribution for whole surface follows Gaussian frequency distribution, local concentration of elements distribution shows spatial localization which causes surface irregularity, anisotropy and heterogeneity. Three dimensional characteristics of the surface are used for better understanding of the joint real behavior and illustrate aperture and contact distribution changes under shear process. The two surfaces are overlapped numerically to form aperture and bring them to contact with each other until a specific distance (aperture obtained from experiment) is achieved. The distance is obtained from initial aperture and normal loading test and also from dilation data at different shear displacements. Aperture distribution is determined and visualized under normal stresses and shear displacements using GIS calculation and visualization tools with element size of 0.2mm×0.2mm. In addition, mean aperture, contact ratio and aperture frequency distribution are determined. Aperture distribution is illustrated using developed three dimensional surface characteristics. Increase in normal stress cause slight increase in contact ratio and decrease in aperture size which causes negative changes in aperture frequency distribution, with homogenous aperture distribution. These results indicate that matching of the surfaces increases with increasing normal stresses and without any shear displacement. Increases in shear displacement causes rapid increase in aperture size and decreases in contact ratio which causes positive changes in aperture frequency distribution, resulted to spatially localization and heterogeneous distribution of the aperture. The surfaces anisotropy is the main reason for the heterogeneity in aperture distribution. Thus, producing un-matching between the surfaces with increasing shear displacement.

# References

Bandis, S.C., Lumsden,A.C., Barton, N.R., (1983): Fundamentals of rock joint deformation, Int. J. of Rock Mech. Min.Sci.&Geomech. Abstr. 20(6):246-268.

Barton, N., (1982): Modeling rock joint behavior from in situ block tests: Implication for nuclear waste repository design. Technocal report ONWI-308, Colombus, Ohaio.

Belem, T., Homand-Etinne,F., Souley, M. (2000): Quantitative parameters for rock joint roughness. Rock Mech. And Rock Engng. 33(4), 217-242.

Brown, S. R., and C. H. Scholtz, (1985): Broad bandwidth study of the topography of natural rock surfaces, J. Geophys. Res., Vol. 90, No.B14:12,575–12,582.

Dijk, P., and B. Berkowitz, (1999): Investigation of flow in Water saturated rock fractures using nuclear magnetic resonance imaging (NMRI), Water Resour. Res., Vol. 35, No: 2, 347–360.

Esaki, T. Ikusada, K. Akikawa A. (1995): Surface roughness and hydraulic properties of sheared rock. Proceeding of Fractured and jointed rock masses, Rotterdam, Balkema: 393-398.

Esaki, T. Shouji, D., Jiang, Y., Wada, Y., Mitani, Y. (1998): Relationship between mechanical and hydraulic aperture during shear – flow coupling test. Proceeding of $10^{th}$ Japan Symposium on Rock Mechanics, 91-96.

Fardin, N. Stephenson, O. and Jing, L. 2001, "The scale dependence of rock joint surface roughness". Int. J. Rock Mech. Sc. Geomech. Abstr. vol. 38: 659-669.

Gale, J. E., (1987): Comparison of coupled fracture deformation and fluid flow models with direct measurements of fracture pore structure and stress-flow properties, Proceedings of the 28th U.S. Symposium on Rock Mechanics, A.A.Balkema:1213–1222.

Gale, J., MacLoad, R., and LeMessurier, P. (1990): Site characterization and validation – measurement of flow rate, solute velocities and aperture variation in natural fractures as a function of normal and shear stress, stage 3. Stripa project. Technical report TR 90-11. SKB, Stockholm.

Gentier, S., (1986): Morphologie et compartment hydrome´canique d'une fracuture naturelle dans un granite sous contrainte normale, Ph.D. thesis, Univ. d'Orle´ans, Orle´ans, France.

Gentier, S., D. Billaux, and L. van Vliet, (1989): Laboratory testing of the voids of a fracture, Int. J. Rock Mech. Rock Eng., Vol. 22: 149–157.

Gentier, S., Riss, J., Archambault, G., Flamand, R. & Hopkins, D. L. 2000. Influence of fracture geometry on sheared behavior *International Journal of Rock Mechanics and Mining Sciences and Geomechanics Abstracts,* 37**:** 161-174.

Grasselli G, Egger P. 2000. 3D surface characterization for the prediction of the shear strength of rough joint. In Proc Eurock 2000, Aachen, Germany. p. 281-6.

Grasselli,G. Egger,P. (2003): Constitutive law for rock joints based on 3-D surface parameters. Int. J. of Rock Mech. Min.Sci.&Geomech. Abstr. 40, 25-40.

Hakami E., Barton, N. (1990): Aperture measurements and flow experiments using transparent replica of rock joints, rock joint proceeding, Leon, Norway.

Hakami E. (1992): joint aperture measurements – an experimental technique, fractured and jointed rock masses proceeding, Lake Tahoe, California.

Hakami E. (1995): Aperture distribution of rock fractures. PhD thesis, Division of Engineering Geology, Department of Civil and Environmental Engineering, Royal Institute of Technology, Stockholm, Sweden.

Accepted in Journal of Rock Mechanics and Rock Engineering- 23 -

Hakami E., and Larsson, E. (1996): Aperture measurements and flow experiments on a single natural fracture. Int. J. Rock Mech. Min. Sci. 33, pp. 395–404.

Hans J., Boulon M. (2003): A new device for investigation the hydro-mechanical properties of rock joints. Int. J. Numer. Anal. Meth. Geomech. 27, pp. 513-548.

Iwano M, Einstein HH. (1993): Stochastic analysis of surface roughness, aperture and flow in a single fracture, EUROCK93, Lisbon, Portugal, pp.135-141.

Iwano M, Einstein HH. (1995): Laboratory experiments on geometric and hydromechanical characteristics of three different fractures in granodiorite. Proceedings of the Eight International Congress on Rock Mechanics, vol. 2. Rotterdam: Balkema: 743–50.

Johns, R. A., (1991): Diffusion and dispersion of solute in a variable aperture fracture, Ph.D. Dissertation, Stanford University, Stanford, CA.

Keller, A., (1998): High resolution, non-destructive measurement and characterization of fracture apertures, Int. J. Rock Mech. Min. Sci. 35, 8, pp. 1037–1050.

Kulatilake, P. H. S. W., Shou, G., Huang, T. H. & Morgan, R. M. 1995. New peak shear strength criteria for anisotropic rock joints *International Journal of Rock Mechanics and Mining Sciences and Geomechanics Abstracts,* 32**:** 673-697.

Kumar, A. T., Majors, A. P. and Rossen, W. (1997): Measurements of aperture and multiphase flow in fractures with NMR imaging, SPE Form. Eval.: 101–107.

Lanaro, F. 2000. A random field model for surface roughness and aperture of rock fractures *International Journal of Rock Mechanics and Mining Sciences and Geomechanics Abstracts,* 37**:** 1195-1210.

Lee and T. F. Cho, (2002): Hydraulic Characteristics of Rough Fractures in Linear Flow under Normal and Shear Load, Rock Mech. Rock Engng. 35 (4), 299–318

Mitani, Y., Esaki, T., Zhou, G., Nakashima, Y. (2003): Experiments and simulation of Shear – Flow Coupling properties of rock joint. Proceeding of 39th Rock mechanics conference: Verlag Gluckauf(GMBH) (pub), USA: 1459-1464.

Mitani, Y., Esaki, T. Sharifzadeh, M., and Vallier, F. (2003): Shear – Flow coupling properties of rock joint and its modeling Geographical Information System (GIS). Proceeding of 10th ISRM conference, South African Institute of mining and metallurgy: 829-832.

Ohtani T., Nakashima , Y. , Nakano, T. , Muraoka H. (2000): X – Ray CT imaging of pores and fractures in the Kakkonda granite, NE japan, Proceedings World geothermal congress , Kyushu – tohoku, Japan:1521-1526.

Pyrak-Nolte, L. J., Myer, L. A., Cook, N. G. and P. A. Witherspoon, (1987): Hydraulic and mechanical properties of natural fractures in low permeability rock, Proceedings of the Sixth International Congress on Rock Mechanics, A. A. Balkema Montreal, Canada: 225–231.

Seidel, J.P.,Haberfielf,C.M. (2002): A theoretical model for rock joints subjected to constant normal stifness and direct shear. Int. J. of Rock Mech. Min.Sci.39:539-553.

Sharifzadeh, M., Mitani, Y., Esaki, T., Urakawa F., (2004a) "An investigation of joint aperture distribution using surface asperities measurement and GIS data processing" Pp165-171, 3rd Asian Rock Mechanics Symposium (ARMS3), Mill press, Kyoto, Japan.

Sharifzadeh, M., Mitani, Y., Esaki, T., Urakawa F.,(2004b) "Determination of evaluation method of rock joint aperture distribution", Pp565-568, EUROCK2004, A.A. Balkema, Salzburg, Austria.

Sharifzadeh, M. (2005) Experimental and Theoretical Research on Hydromechanical coupling properties of rock joint, PhD thesis, P226,Kyushu University, Japan.

Stephanie, P. Bertels, David A. DiCarlo, and Martin J. Blunt, (2001): Measurement of aperture distribution, capillary pressure, relative permeability, and in situ saturation in a rock fracture using computed tomography scanning, Water Resour. Res., VOL. 37, NO.3:649–662.

Tse R. & Cruden D. M., 1979, Int. J. Rock Mech. Min. Sci. & Geomech. Abstr. Vol. 16, pp 303-307.

Xie, H. P., Wang, J. A. & Xie, W. H. 1997. Fractal effects of surface roughness on the mechanical behavior of rock joints Chaos Solutions and Fractals, 8: 221-252.

**Figure captions:**

Fig.1. Procedure for determination of aperture distribution.

Fig. 2.Schematic view of designed laser scanner and modified table for joint's surfaces measurement and calibration test results (scale not respected).

Fig. 3. Accuracy test for table stopping time.

Fig. 4. Errors originated from specimen installation on laser scanner table

Fig. 5. Selection of real surface data from measured data

Fig. 6.Joint lower and upper surfaces asperity's heights, slope angle, and aspect distribution.

Fig. 7. Upper and lower surfaces three dimensional view along with profiles in x- and y- direction to show surface roughness irregularity and elements concentration on local areas over the joint's surfaces

Fig. 8. Procedure for joint Initial aperture determination from normal loading test.

Fig. 9. Joint closure determination procedure for joint initial aperture calculation under different normal stresses.

Fig. 10. Shear test results under 3MPa of normal stress, showing the changes of shear stress and normal displacement (dilation) versus shear displacement

Fig. 11. Aperture distribution map (left) and frequency distribution (histogram-right) with mean aperture and percent of contact ratio under different normal stresses.

Fig. 12. Aperture distribution map (left) and aperture frequency distribution (histogram-right) with mean aperture and percent of contact ratio at different shear displacements (under 3MPa of normal stress).

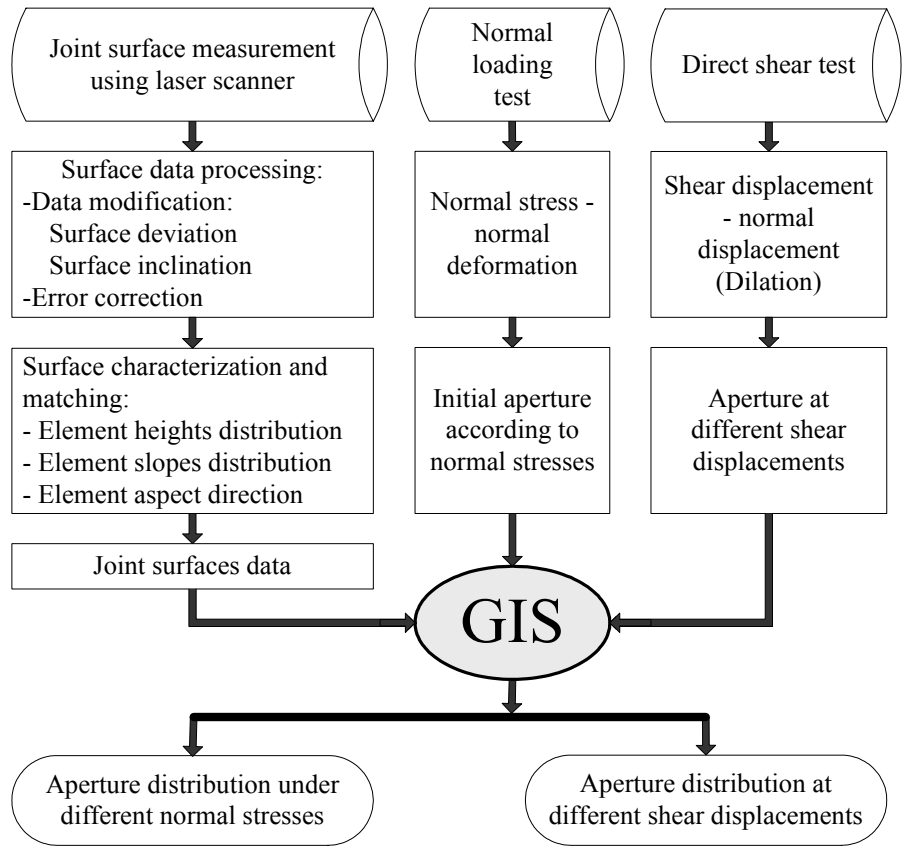

Fig. 1. Procedure for determination of aperture distribution.

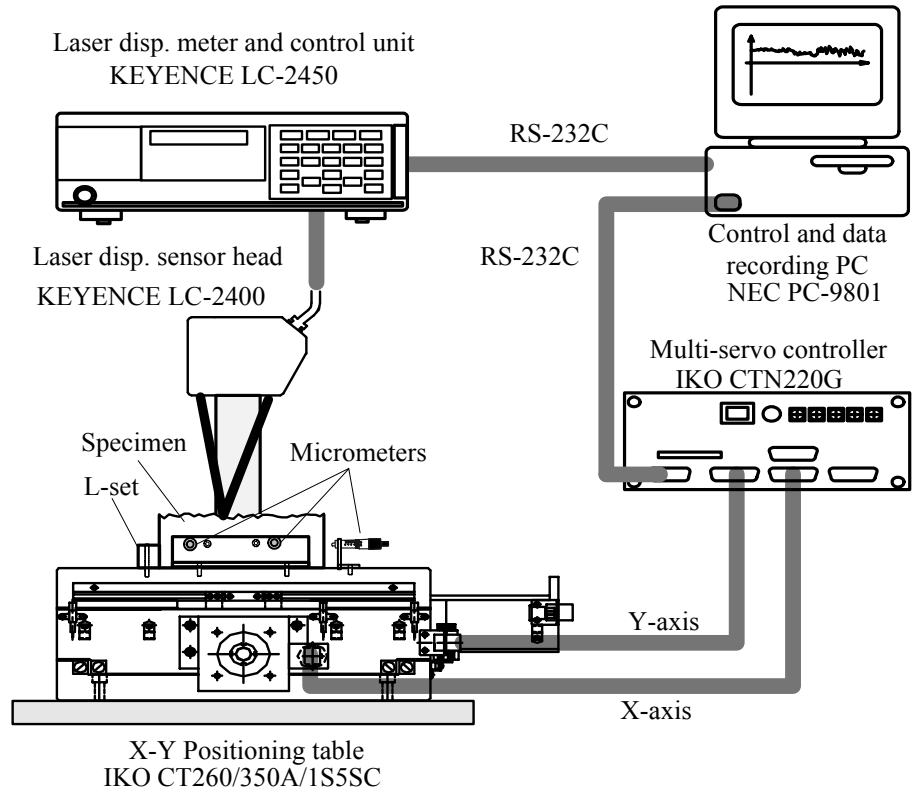

a) Schematic view of three dimensional laser scanner system

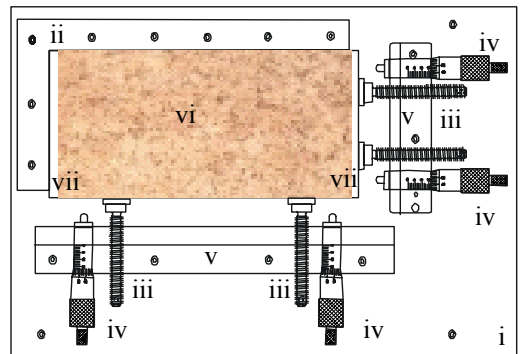

i) Modification tools base plate, ii) L-set, iii) Fixing screw pins, iv) Micrometers, v) L-shape holders, vi) Specimen, vii) Steel plates

b) Plan view of improved laser scanner table

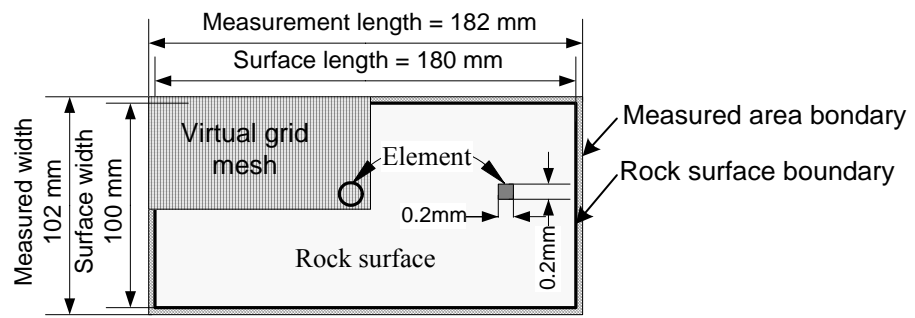

c) Schematic view of surface measurement on virtual grid elemnt

Fig. 2. Schematic view of designed laser scanner system and modified table for joint surfaces measurement and calibration test results (not to scale).

- 27 -

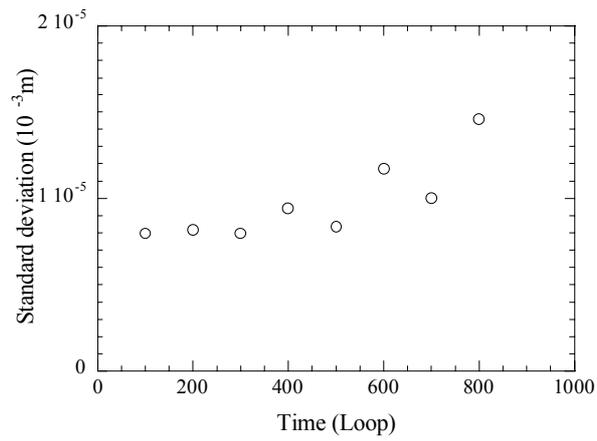

Fig. 3. Accuracy test for table speed

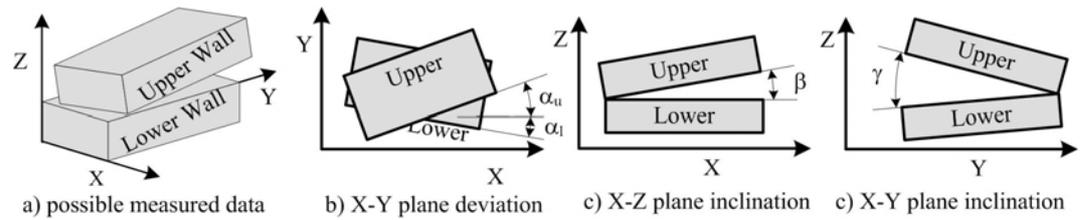

Fig. 4. Errors originated from specimen installation on laser scanner table

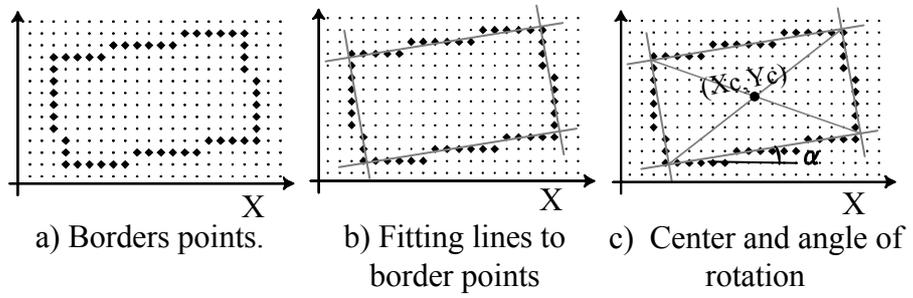

a) Borders points.   b) Fitting lines to border points   c) Center and angle of rotation

Fig. 5. Selection of real surface data from measured data

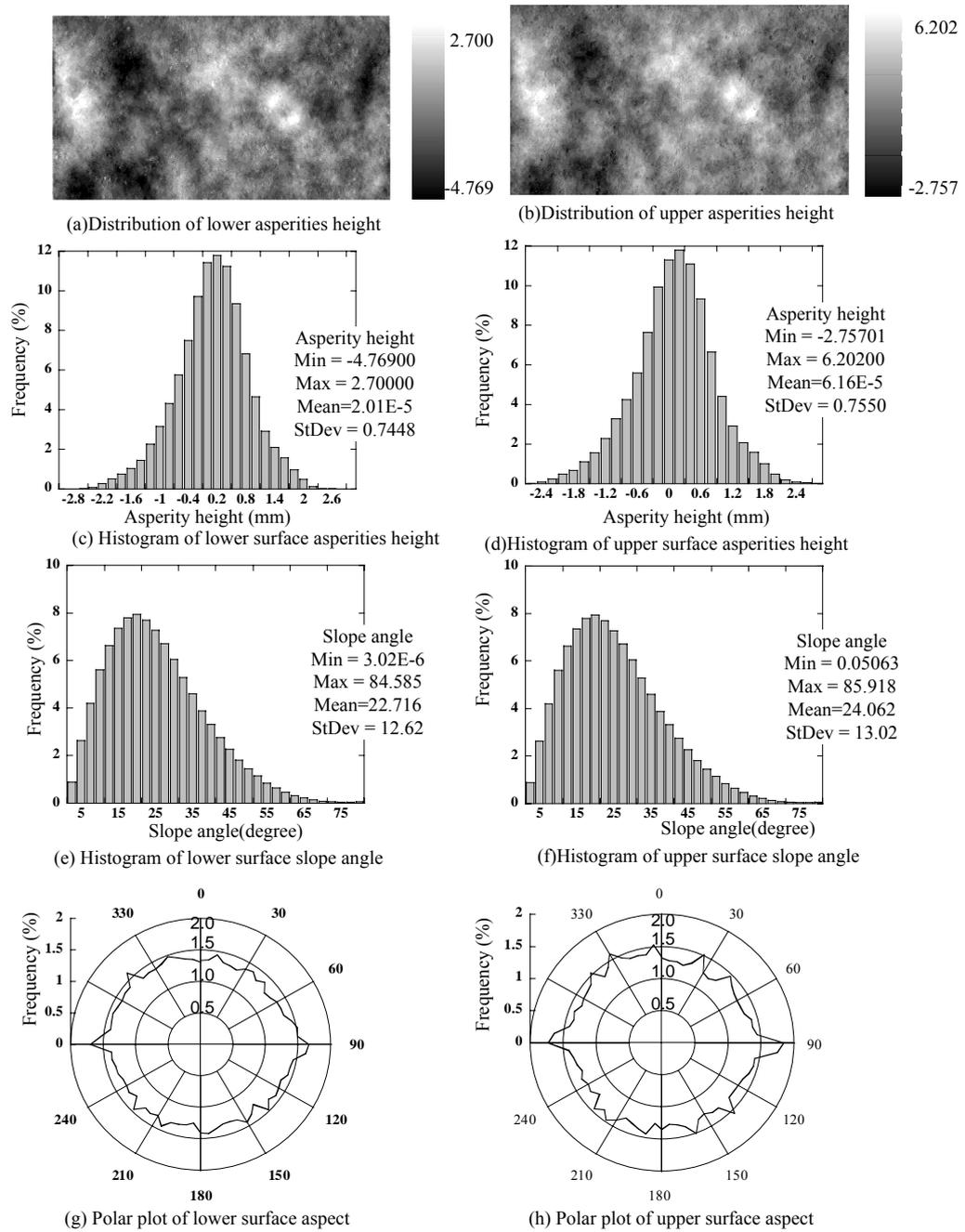

Fig. 6. Joint lower and upper surfaces asperities height, slope angle, and aspect distribution.

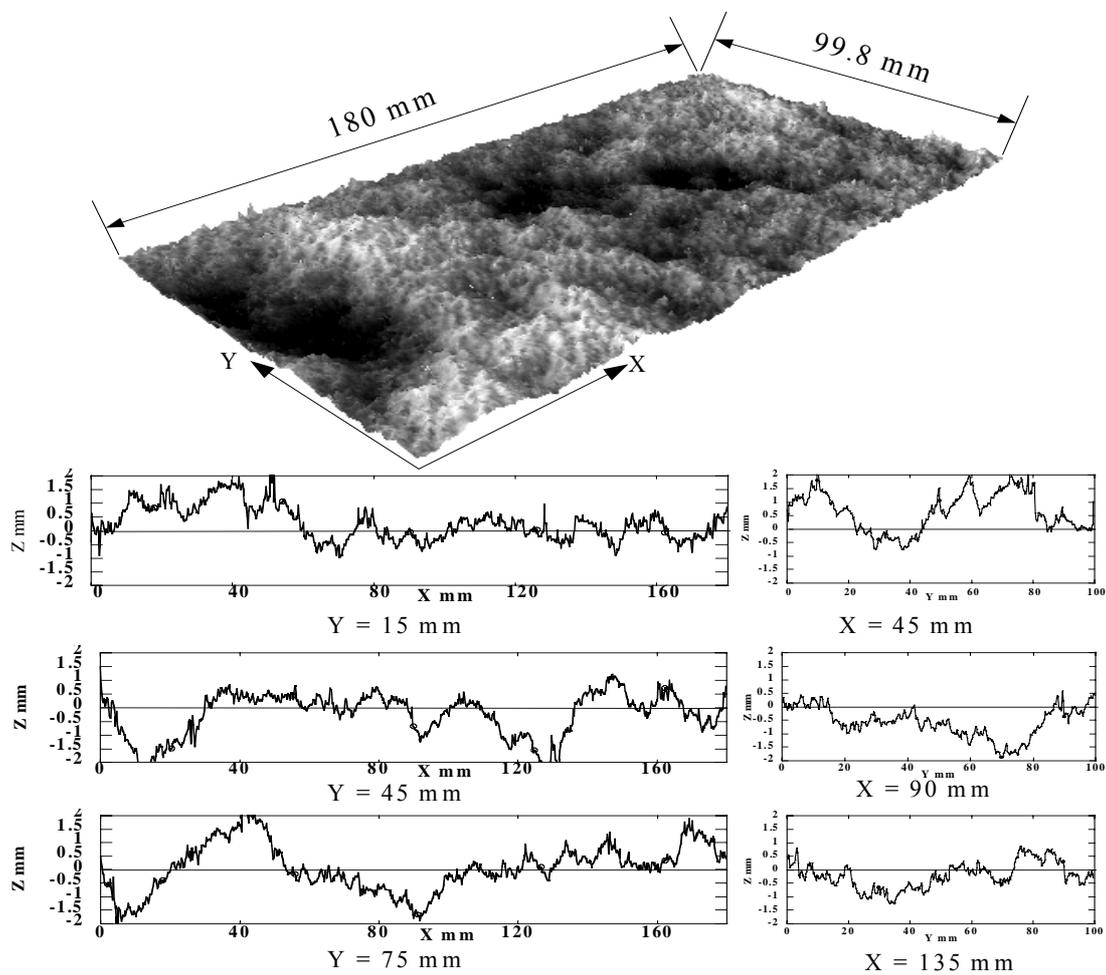

a) Three dimensional view and Profiles in x- and y- direction on upper surface

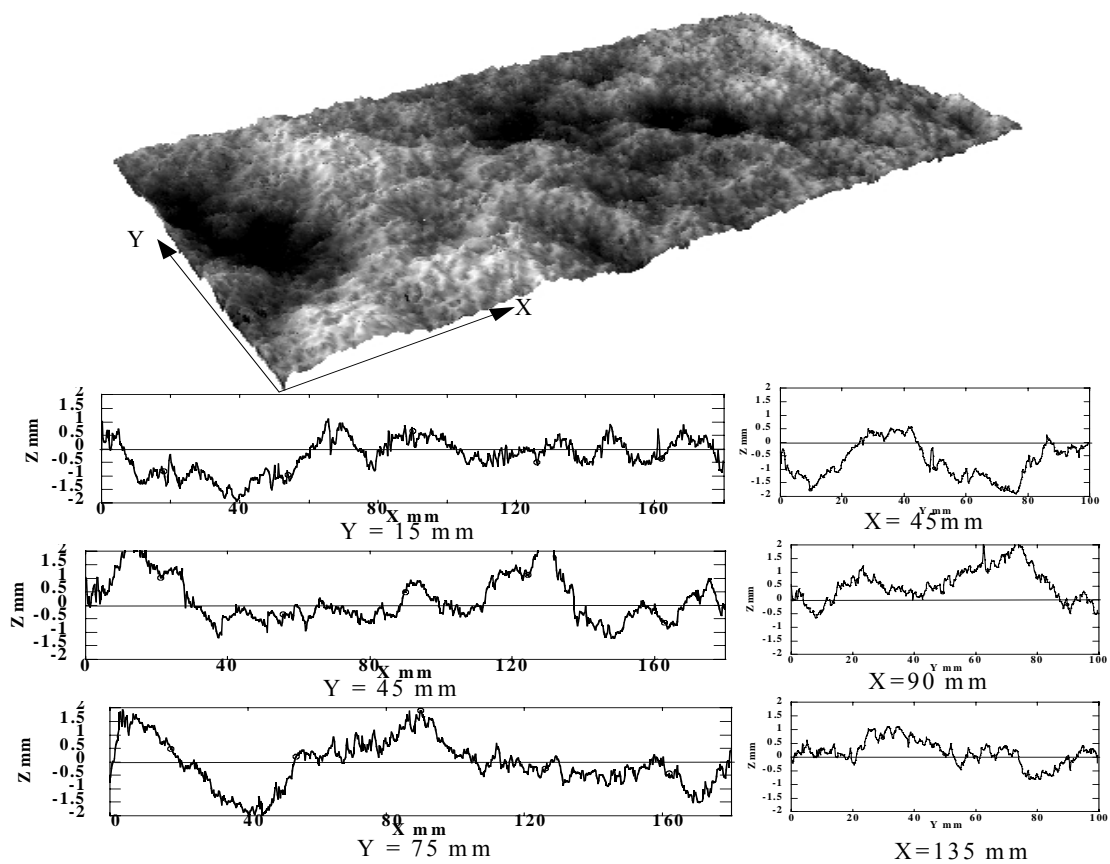

b) Three dimensional view and Profiles in x- and y- direction on lower surface

Fig. 7. Upper and lower surfaces three dimensional view along with profiles in x- and y- direction to show surface roughness irregularity and elements concentration on local areas over the joint surfaces.

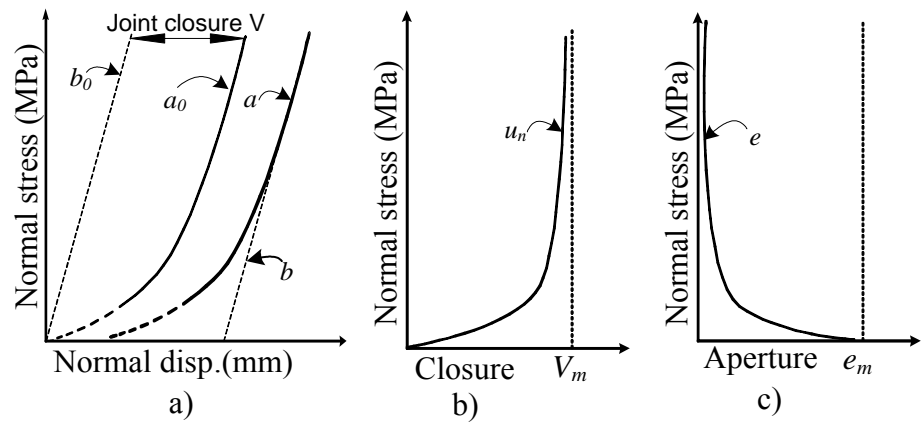

Fig.8. Procedure for joint Initial aperture determination from normal loading test.

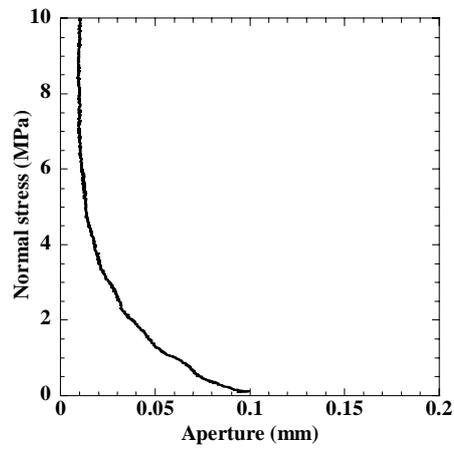

Fig. 9. Initial aperture versus normal stress for study case

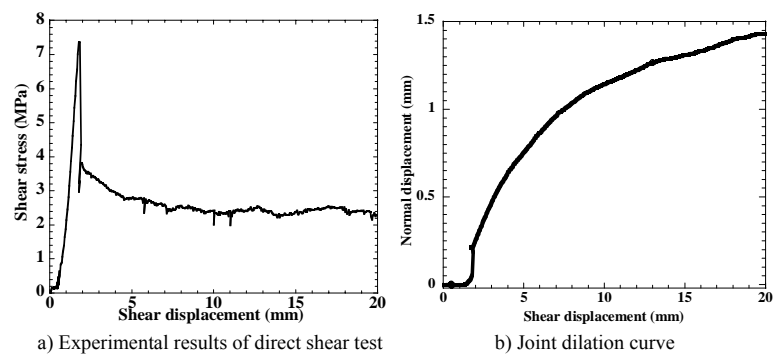

a) Experimental results of direct shear test    b) Joint dilation curve

Fig. 10. Shear test results under 3MPa of normal stress, showing the changes of shear stress and normal displacement (dilation) versus shear displacement

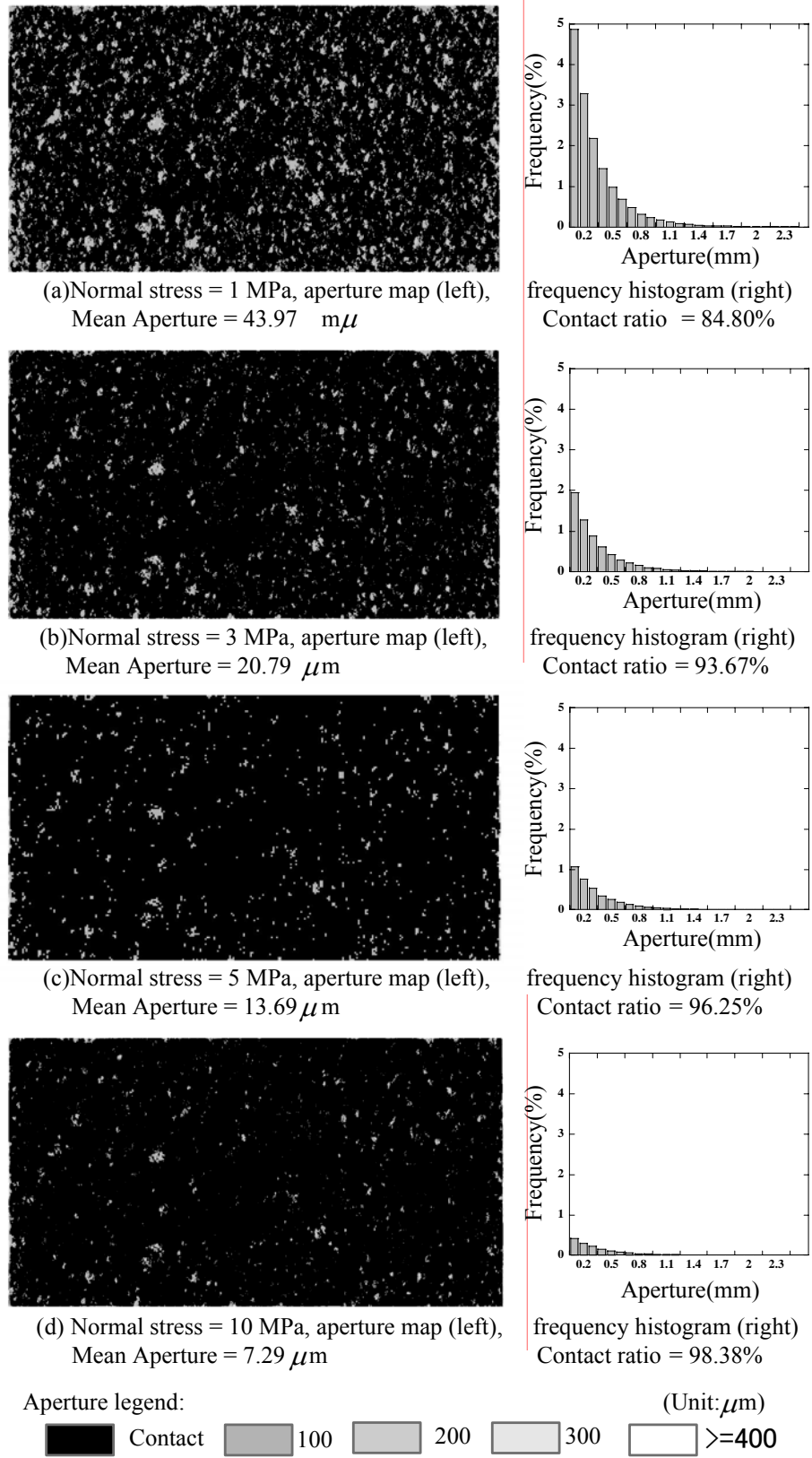

Fig. 11. Aperture distribution map (left) and frequency distribution (histogram-right) with mean aperture and percent of contact ratio under different normal stresses.

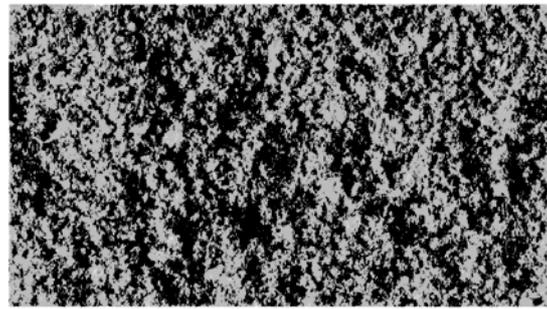 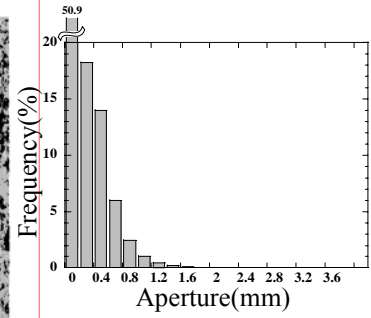

(a) Shear displacement 1 mm, aperture map (left), frequency histogram (right)
Mean Aperture = 131.19 $\mu$m   Contact Ratio = 50.8%

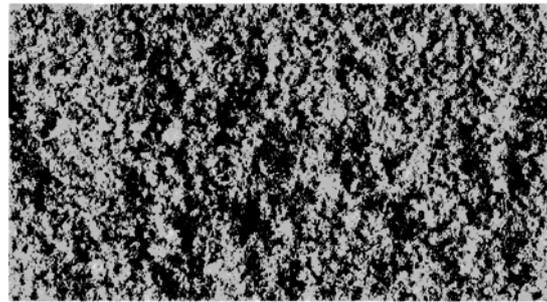 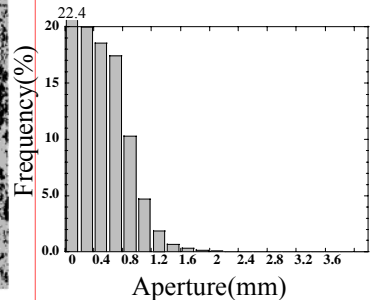

(b) Shear displacement 2 mm, aperture map (left), frequency histogram (right)
Mean Aperture = 325.13 $\mu$m   Contact Ratio = 22.4%

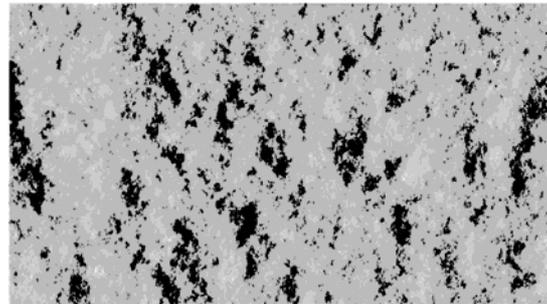 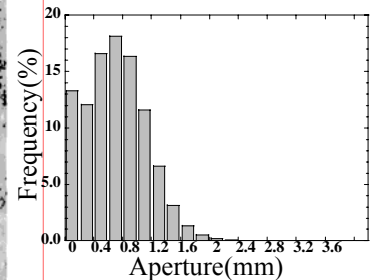

(c) Shear displacement 3 mm, aperture map (left), frequency histogram (right)
Mean Aperture = 520.76 $\mu$m   Contact Ratio = 13.3%

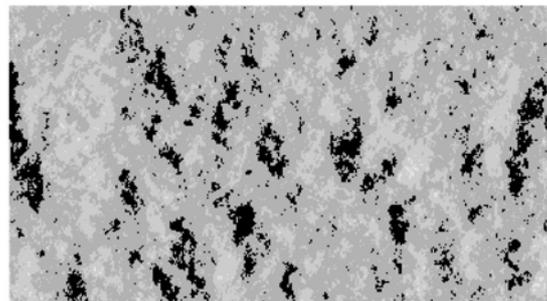 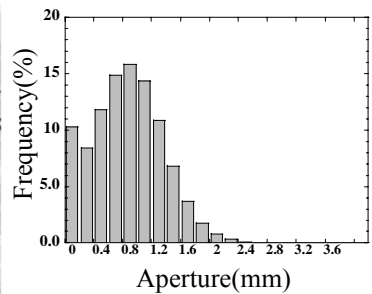

(d) Shear displacement 4 mm, aperture map (left), frequency histogram (right)
Mean Aperture = 677.78 $\mu$m   Contact Ratio = 10.3%

Figure 13 continued

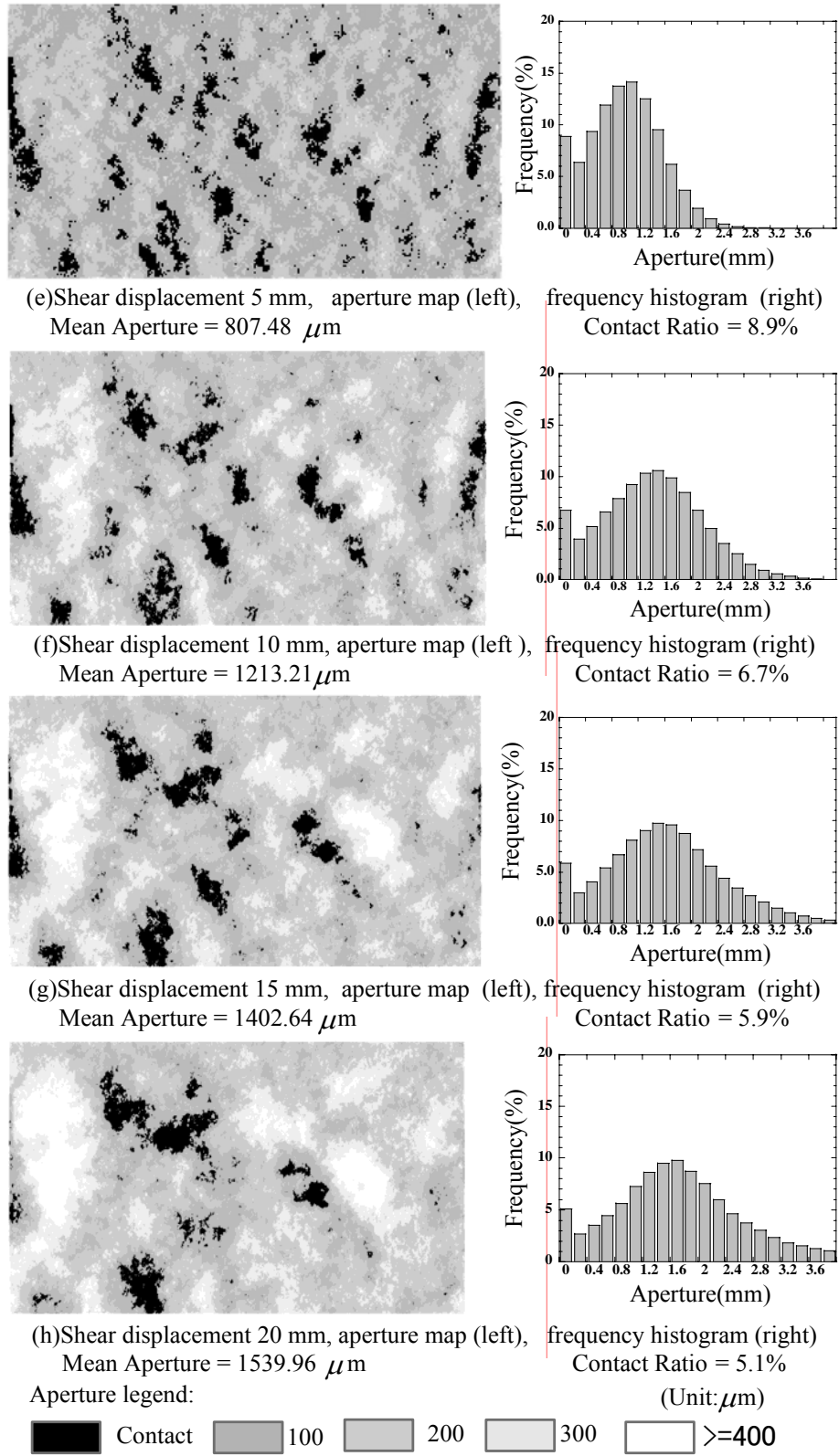

(e) Shear displacement 5 mm, aperture map (left), frequency histogram (right)
Mean Aperture = 807.48 μm    Contact Ratio = 8.9%

(f) Shear displacement 10 mm, aperture map (left), frequency histogram (right)
Mean Aperture = 1213.21 μm    Contact Ratio = 6.7%

(g) Shear displacement 15 mm, aperture map (left), frequency histogram (right)
Mean Aperture = 1402.64 μm    Contact Ratio = 5.9%

(h) Shear displacement 20 mm, aperture map (left), frequency histogram (right)
Mean Aperture = 1539.96 μm    Contact Ratio = 5.1%

Aperture legend:    (Unit: μm)

Contact   100   200   300   >=400

Fig. 12. Aperture distribution map (left) and aperture frequency distribution (histogram-right) with mean aperture and percent of contact ratio at different shear displacements (under 3MPa of normal stress).